# Secure Web Access Control Algorithm

**Filip Ioan, Szeidert Iosif, Vasar Cristian,**

Department of Control System Engineering,

"Politehnica" University of Timisoara, Faculty of Automation and Computers

300223 Timisoara, Bv. V. Parvan, No.2, ROMANIA

ifilip@aut.utt.ro

*Abstract: The paper presents a flexible and efficient method to secure the access to a Web site implemented in PHP script language. The algorithm is based on the PHP's session mechanism. The proposed method is a general one and offers the possibility to implement a PHP based secured access to a Web site, through a portal page and using an additional script included in any site's page, which is required to be accessed only by registered users. This paper presents the design, implementation and integration of the algorithm on any generic WEB site.*

*Keywords: PHP, Web Site, Secure Access, Session*

## 1. Introduction

The overwhelming Internet network expansion and specially of Web site application's, lead to the imminent security issues on the access to specific data available to a restricted set of users. Therefore, currently the informational attack on secured data resources – available only based on access accounts (username and password) - shows a huge increase. From this point of view, the exposed thematic of this paper presents a large interest.

The high usage rate of PHP language in relation with the classical HTML language (and eventually the MySQL database server) in the WEB applications implementation is promoted by a continuous development of this scripting language, which has presently reached the 4.3 version [1]. The proposed solution is fully functional on PHP versions starting with 4.1.1. The basic idea of the algorithm is to provide a secured portal page, which allows only an authorized access, to the whole Web site, and deny direct access by jumping over the portal page. There is a script included on the beginning of each page of the Web site, having a watchdog role that blocks any unauthorized access to the available data from the site's pages. In the paper there are presented the source codes of the secured portal page (the *enter.php* script) and also the code of the PHP script that secures the access to any page of the site (the *guardian.php* script). The described solution denies any direct access (without passing through the portal page) to any WEB site's page that includes the guardian script at the beginning. The main idea assumes the existence of an opened PHP session during the entire browsing process on the site's pages performed by a valid user and a permanent checking of a session variable's existence on each in/out browsing between pages (both in the case of browsing the website forward and backward) [8].

## 2   Design and implementation of the secure solution

The informational diagram of the securing mechanism is presented in figure 1. Also, for a better understanding of the below presented details, in tables 1 and 2 there is presented the entire source code of the application. As it has been mentioned, the proposed solution involves the existence of a portal file for an authorized access (table 1 – *enter.php* script), respectively of a script file (table 2 – *guardian.php* script) that has to be included on each site's page that is necessary to be secured. By the term 'securing' it is understood that the access to any site's page (excepting the portal page) can be done only using a valid username and password [2][6].

When a user requests an access to the site's resources via the portal page, the first PHP section code (figure 1 – PHP code 1) opens a PHP session that remains active until the client's browser is closed. Afterwards, it is verified that the $id variable is set on the 'set' value. Normally, at the first access of the portal page, the variable is not set, as a result a jump is performed (trace 1 – figure 1) over the entire script's PHP code to a HTML sequence code (a HTML form), which requests an authentication by introducing, from the keyboard, the username with the corresponding password [3]. After this input operation is done, the script is automatically recalled (trace 2). The recall of the script is executed with passing of the username, password and *$id* variable values as input parameters.

The script checks if the *$id* variable is set on '*set*' value (the condition is fulfilled at the second call) and begins the execution of the second PHP sequence code (PHP code 2). In this PHP sequence, the validity of the username and password is checked. In the presented example, there has been considered that the authentication data of the valid users are stored into a table, within a MySQL database. It is obviously that those authentication data can be stored also into another type of resources, such as: simple text files, tables from different database types, etc. In all those cases, the password is kept encrypted (using the *md5* encrypting algorithm implemented by *md5( )* PHP function)[4][5]. For that reason, in this code sequence the typed password is initially encrypted and afterwards compared with his encrypted form stored in the MySQL table. After a query of the

table is performed, if no user is found with the specified username and password (no record found), the authentication process fails and the HTML sequence code is recalled (restarting another authentication attempt by requesting another username and password – trace 3), and signaling the failure of the previous access through a corresponding error message. So the algorithm is restarted.

If the user is valid, then during the PHP session a session variable is assigned with the valid username. (*$_SESSION["user"]=$user*).

Tabel 1 *Enter.php* script

```
<?PHP
// session start
session_start();
$eroare = "";
// MySQL server connects
$conectare = @mysql_connect("localhost", "root", "") or die("Error");
$bazadate = @mysql_select_db ("test");
// id="set" – only for the second access
if(isset($_POST["id"]) && $_POST["id"] == "set") {
        $username = $_POST["name"];
// encrypt the password
        $parola = md5($_POST["parole"]);
        if($eroare == ""){
// query the table
$sql = "SELECT name, parole FROM table1
WHERE name = '$username' AND parole = '$parola' ";
$query_result = mysql_query ($sql);
$num_of_rows = mysql_num_rows ($query_result);
                if($num_of_rows == 1) {
// valid user
                        $_SESSION["user"] = $username;
                        mysql_close($conectare);
// redirect to the next page
                header("Location: page1.php");
                }
// invalid user
                if($num_of_rows == 0) {
                        $eroare = "User unregistered!";
                        mysql_close($conectare);
                }
        }
  }
else {
// reset variables
        $username = "";
        $parola = "";
```

```
}
// end of PHP scrpt – begin the HTML form
?>
<form name="intrare" method="post" action="enter.php">
<?PHP echo $eroare; ?>
    <input name="id" type="hidden"  value="set">
  Name: <input name="name" type="text" size="20" value="<?PHP echo
$username; ?>"><br>
  Password:<input name="parole" type="password" size="20" maxlength="20"
value="">
    <input type="submit" name="nsubmit" value="LOGIN">
</form>
```

In the last phase a redirection to the first page with secured data is accomplished (page1.php script) and the access to the secure area site is granted. In this page, the first PHP code line must be: *include("guardian.php")*. The included script file (table 2) maintains the PHP session open (*session_start( )*) and checks if the session variable *$_SESSION["user"]* is set.

Table 2 *Guardian.php* script

```
<?PHP
// session start
session_start();
// is set session variable?
if (!isset($_SESSION["user"]))
// recall the enter page
{
         header("Location: enter.php");
exit;
}
?>
```

If the session variable is set, the access was requested by a valid user (trace 2). Also, the included watchdog procedure is terminated (*exit* command) and the next line (after the *include* command) from the secured page is executed. Otherwise, it means that an unauthorized direct access is in progress and there is conducted a redirecting to the portal page and to the authentication procedure (trace 3'). As a result, the whole algorithm is restarted. In this order of idea, any other Web site's page that includes the *guardian.php* script is secured. So, any direct access (skipping of the portal page) is rejected and automatically redirected to the portal page (*enter.php* script file). The session module cannot guarantee that the information stored in a session is only viewed by the user who created the session. So must to take additional measures to actively protect the integrity of the session, depending on the value associated with it [1][8].

Taking into consideration the above stated facts, only the user name is stored, without the corresponding password, into a session variable (and also in the session file). However, in order to make viable the securing mechanism this data are enough.

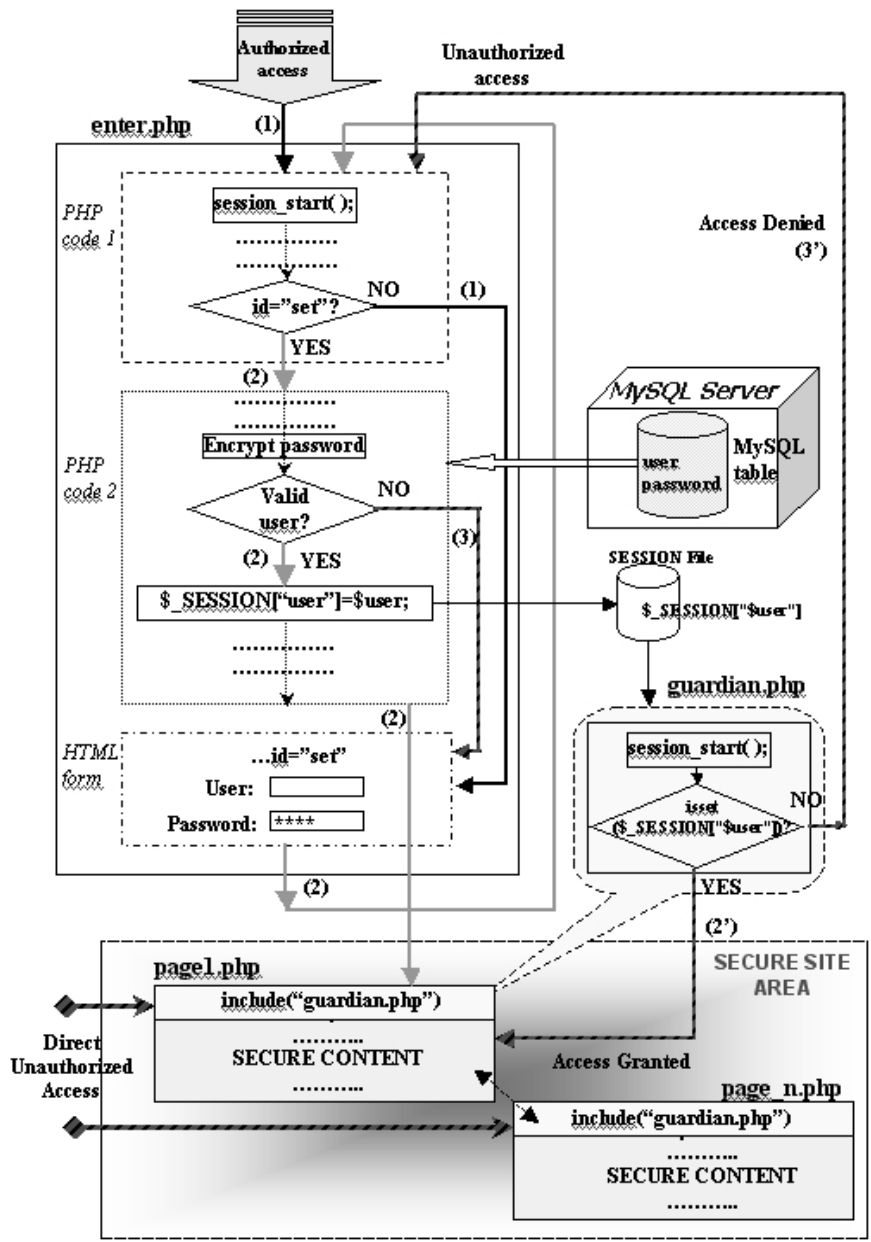

Fig.1 Informational transfer on the securing mechanism

## 3  Integrating into a Web site

As it has been already stated, the modularization of the application allows a facile integration into any WEB site that needs to be secured. However, there are some preliminary steps that need to be performed.

Firstly, due to the fact that in the presented example (table 1), the data related to the users are stored in a MySQL table (user accounts: username and password), the 'table1' table form 'test' database, this table must be previously created and must contain at least one user. Therefore, in MySQL must be executed the following operations.

> USE test;

>CREATE TABLE table1 (name varchar(20), parole varchar(60));

In the next step, assuming that it is desired the creation of an account for a user having the username '*ion*' and the password '*parola*', it is performed by encrypting the password running a single line PHP script [4]:

<? PHP echo md5('parola');?>

On the screen will be displayed the encrypted password:

8287458823facb8ff918dbfabcd22ccb

It can be noticed the larger size of the encrypted password's form, this being the reason for the larger size in the definition of the '*parole*' field.

The next step, is to execute in MySQL the following command that creates the desired user account:

> INSERT INTO table1 VALUES('ion', '8287458823facb8ff918dbfabcd22ccb');

In this moment, the user account has been created. Assuming that the aimed WEB site has the pages 'page1.php', 'page2.php', etc., each one of the mentioned scripts must include firstly the following code line:

```
<?PHP include("guardian.php"); ?>
```

The access to WEB site's pages can be fulfilled in this case only after a successful access of the entry portal. Any direct access (without the passage of the entry portal) to any Web site's pages is rejected and rerouted to the entry portal page (the enter.php script file).

**Conclusions**

The proposed solution is relatively simple, very secure and can be easily implemented by any PHP language programmer. The modularized approach (using two PHP scripts) provides a facile usage for the securing of any PHP language based WEB site implementation. It is very easy to use this method to secure even existing Web sites, due to the minor changes requited on each page of the site. The basic idea that leads to the implementation of this algorithm was the session mechanism powered by PHP. It is mentioned that starting with the 4.1.1. PHP version, the operating procedure of the session mechanism has been modified, so that the proposed solution is not operable on older PHP versions [8]. However, with some slightly modifications, it can be adapted also for the case of some older PHP versions.